\documentclass[preprint,12pt]{elsarticle}

\usepackage{amsmath,amssymb,amsfonts}
\usepackage{braket}
\usepackage{algorithmic}
\usepackage{graphicx}
\usepackage{textcomp}
\usepackage{xcolor}
\usepackage{caption}
\usepackage{float}
\usepackage[unicode=true, psdextra]{hyperref}
\usepackage[numbers]{natbib}

\begin{document}

\begin{frontmatter}



\title{Quantum Information Analysis in a q-Deformed Deng--Fan Model}

\author[thai,may1]{P. O. Amadi}
\author[may2,cross]{Collins O. Edet}
\author[br1,br2]{A. R. P. Moreira}
\author[thai]{P. Kalasuwan}
	\ead{pruet.k@psu.ac.th}
\author[may2,may3]{N. Ali}
	\ead{norshamsuri@unimap.edu.my}
\author[may1,may3]{R. Endut}
\author[may1,may3]{S. A. Aljunid}


\affiliation[thai]{
		organization={Division of Physical Science, Faculty of Science},
		addressline={Prince of Songkla University}, 
		city={Hat Yai},
		postcode={90110}, 
		state={Songkla},
		country={Thailand}}
\affiliation[may1]{
		organization={Faculty of Intelligent Computing, Universiti Malaysia Perlis},
        postcode={02600}, 
		city={Arau Perlis},
		country={Malaysia}}
\affiliation[may2]{
		organization={Faculty of Electronic Engineering \& Technology, Universiti Malaysia Perlis},
        postcode={02600}, 
		city={Arau Perlis},
		country={Malaysia}}
 
\affiliation[cross]{
		organization={Department of Physics, University of Cross River State,},
		state={Calabar},
		country={Nigeria}} 
\affiliation[br1]{
		organization={Secretaria da Educa\c{c}\~{a}o do Cear\'{a} (SEDUC),},
		addressline={ Coordenadoria Regional de Desenvolvimento da Educa\c{c}\~{a}o (CREDE 9)},
        postcode={62880-384}, 
		city={Horizonte},
		state={Cear\'{a}},
		country={Brazil}}   
\affiliation[br2]{
		organization={Postgraduate Program in Electrical and Computer Engineering, Federal University of Cear\'{a}},
		addressline={ Coordenadoria Regional de Desenvolvimento da Educa\c{c}\~{a}o (CREDE 9)},
        postcode={62010-560}, 
		city={Sobral},
		state={Cear\'{a}},
		country={Brazil}}         
\affiliation[may3]{
		organization={Centre of Excellence Advanced Communication Engineering (ACE), Universiti Malaysia Perlis (UniMAP)},
		city={Arau},
		postcode={02600}, 
		state={Perlis},
		country={Malaysia}}

\cortext[1]{Corresponding author}

\begin{abstract}
We introduce a $q$-deformed Deng-Fan potential ($q$DFP) model that enables controlled modulation of short-range repulsion and long-range attraction while preserving the equilibrium configuration. The model is solved exactly within the framework of the time-independent Schr\"odinger equation, yielding closed-form expressions for the energy eigenvalues and wave functions in terms of hypergeometric functions. We show that the deformation parameter $q$ induces non-uniform spectral shifts and a redistribution of bound states. In particular, for $q<1$, the system exhibits spectral compression and enhanced spatial localization. In addition, we investigate the system from an information-theoretic perspective using Shannon entropy, Fisher information, and Fisher--Shannon complexity measures in both position and momentum spaces. The results reveal that the deformation parameter governs the redistribution of quantum information, establishing a direct connection between spatial confinement and momentum delocalization in accordance with the Bia{\l}ynicki-Birula and Mycielski entropic uncertainty principle. Stronger deformation pushes the quantum state further from the minimum-uncertainty configuration, increasing the entropic excess above the BBM bound and reducing the information content about complementary observables, even as position-space localization sharpens. The analysis of entropic and Fisher information densities further shows how the deformation reshapes both the local information content and the structural complexity of the quantum states. We show that in the limit $q\to 1$, the $q$DFP model is reduced to the standard Deng-Fan potential.
\end{abstract}



\begin{keyword}
Deng Fan Potential \sep Schrodinger equation \sep localization  \sep Fisher Information \sep localization \sep  Shannon Entropy \sep Complexity Measures



\end{keyword}

\end{frontmatter}


\section{Introduction}
\label{sec1}

The study of empirical potential functions remains a central theme in molecular physics and quantum chemistry. It provides the necessary framework that describes the energy-distance relation during formation and dissociation of chemical bonds~\cite{hajigeorgiou2010extended}. Among these models, one of the earliest,  proposed by Deng and Fan in 1957~\cite{deng1957potential}, is sometimes referred to as the generalized Morse potential. The Deng Fan potential (DFP) model is a well-known empirical potential that accurately describes the diatomic molecule (DM) energy spectrum and electromagnetic transitions.  It is defined as
\begin{eqnarray}\label{dengfan}
V(x) =\mathcal{D}_e\left(1-\frac{e^{\alpha x_e}-1}{e^{\alpha x}-1}\right)^2,
\end{eqnarray}
where $\mathcal{D}_e$ is the potential depth, $x_e$ is the equilibrium bond length, $\alpha$ is the range of the potential well~\cite{deng1957potential,mesa1998generalized}.

The introduction of the DFP model was basically to address the mathematical shortcomings of the Morse potential~\cite{morse1929diatomic}. While the Morse model is known for its mathematical tractability and inclusion of anharmonicity, it possesses an incorrect boundary condition at the origin. This means that the Morse potential maintains a finite value as the internuclear distance approaches zero~\cite{rong2003comparison, muser2022interatomic}. However, the DFP was proposed to correctly exhibit an infinite repulsive barrier  ($V\to \infty$ as $r \to 0)$  and approaches the dissociation limit $\mathcal{D}_e$ as $r \to \infty$. Unlike conventional Dunham-type analysis~\cite{dunham1932energy}, like the Morse potential, which is restricted to low-order coefficients, the DFP model can incorporate higher-order terms. This allows it to effectively represent the molecular potential over a broad range of vibrational levels, including those near dissociation. This characteristic makes the DFP model an excellent internuclear potential for describing diatomic energy spectra and electromagnetic transitions~\cite{vogt2018accuracy}.

The Deng-Fan potential (DFP) of the family of exactly solvable molecular potentials, and physically equivalent to the Manning-Rosen and Schiöberg potentials~\cite{wang2012equivalence}. Its analytical tractability has enabled DFP extensions, such as the shifted Deng-Fan potential for improved local harmonic behavior and accurate treatment of vibrational and rotational states~\cite{alici2025brational,oyewumi2013thermodynamic}. Approximate and exact solutions of DFP have also been reported in literature~\cite{dong2008arbitary,greene1976variation,oyewumi2012bound,zhang2011approximate}. The DFP model has also been generalized to fractional and higher-dimensional quantum systems, relativistic Klein-Gordon and Dirac equations, and the Dunkl derivative formalism, with applications extending to to information-theoretic analyses, thermodynamic properties, and even DNA denaturation studies~\cite{abu-shady2022general,hassanabadi2012deng,ikot2013solution,halder2025information,onate2018effect,nyeo2001phase,hamzavi2012equivalence,edetikot2022superstatistics,ghanbari2025theoretical,lumb2016rovibrational,oluwadare2018energy}. Despite these advancements, DFP still retains the conventional exponential structure where its short-range repulsive wall and long-range attractive tail are strictly coupled to their intrinsic parameters $\mathcal{D}_e,\alpha$, and $x_e$ and fixed, with no room to tune the potential shape independently of these molecular constants.

The introduction of a deformation parameter $q$ as an effective phenomenological also quantity extends the flexibility of the Deng-Fan potential without sacrificing exact solvability~\cite{angelova2004revisiting, hassanabadi2017deformed, boumali2018statistical}. In realistic molecular systems, it is obvious that empirical pair potentials cannot fully account for deviations arising from anharmonicity, electronic screening, and many-body interactions. Rather than introducing several additional molecular constants, a single deformation parameter is presented which modifies the depth and profile of the interaction potential while preserving the equilibrium bond length. As consequence, the molecular geometry remains unchanged, whereas the effective interaction and vibrational spectrum become continuously tunable through a single parameter. Although $q$ is not interpreted as a fundamental molecular constant, its value maybe inferred by fitting spectroscopic observables, thereby providing a physically meaningful description of deviations from the standard Deng-Fan interaction.  Our proposed $q$DFP therefore offers a flexible analytical framework for investigating a broader class of molecular interaction profiles within the same exactly solvable formalism.

Motivated by this, we propose an exactly solvable $q-$ deformed Deng-Fan potential ($q$DFP), which introduces a single deformation parameter $q \in(0,1]$. This range is physically motivated: at $q=1$, the standard DFP is fully recovered. The deformation parameter $q$ introduces a non-uniform modification of the potential, where the short-range repulsion and long-range attraction respond differently to the deformation. $q<1$ progressively steepens the repulsive core and weakens the long-range attractive tail, which mimick the effect of enhanced confinement without altering the equilibrium bond length $x_e$. For values $q>1$, are excluded; they shift the effective singularity of the denominator $e^{\alpha x}-q$, which produces a non-physical potential profiles that violate molecular admissibility. Our $q$DFP model provides a mechanism for tuning the spectral distribution and spatial localization without changing the equilibrium configuration. We also show that the proposed model satisfies molecular admissibility under the Vashini conditions. Studies on q-deformations provide effective descriptions in a variety of quantum systems, including quantum algebras~\cite{yan1990deformed,schmidt2006deformed}, nuclear systems~\cite{sviratcheva2004physical}, and anharmonic molecular vibrations~\cite{boumali2018statistical, edet2026controllable}. While a first-principles derivation of the deformation parameter is beyond the scope of the present work, $q$ can be calibrated against experimental spectroscopic data for specific molecules. Within our proposed $q$DFP model, the deformation modifies the interaction profile while preserving the equilibrium bond length, thereby providing additional flexibility for describing deviations from the ideal Deng-Fan interaction arising from higher-order anharmonicity ~\cite{angelova2004revisiting}. Varying $q$ therefore offers a systematic means of investigating how changes in the effective interaction influence the vibrational spectrum and the localization properties of molecular wavefunctions.

Again, while the energy spectrum determines the allowed vibrational levels of a molecule, it does not characterize uniquely the spatial structure of the corresponding quantum states. Since the proposed q-deformation primarily modifies the interaction profile rather than simply shifting the eigenvalues, its physical consequences are most naturally reflected in the wavefunctions. For this reason, Shannon entropy, Fisher information, and Fisher-Shannon complexity are employed as complementary descriptors of the molecular bound states. Shannon entropy quantifies the global spatial delocalization of the probability density\cite{shannon1948mathematical,moreira2025testing,
moreira2026quantuminformation}, Fisher information measures its local gradients and sensitivity to structural changes~\cite{fisher1925theory,frieden1992fisher}, while Fisher-Shannon complexity combines these global and local characteristics into a single measure of the structural organization of the quantum state~\cite{vignat2003analysis,romera2004fisher,inyang2025quantum}. Therefore, these information theoretic quantities provide the sufficient characterization of the molecular wavefunction and reveal how the $q$-deformation modifies localization in ways that cannot be inferred from the energy spectrum alone. The Fisher-Shannon complexity measure has been applied to electronic correlations, and the internal organization of quantum systems, to characterize the interplay between disorder and structure of a quantum system~\cite{sen2007fisher, angulo2008atomic,lopez2011statistical}.

The remainder of this paper is organized as follows. In Section~\ref{model}, we introduce the $q$DFP model and discuss its main structural properties. In Section~\ref{exact}, we derive the exact analytical solutions of the corresponding Schr\"odinger equation. 
In Section~\ref{quantum}, we investigate the quantum information aspects of the system through Shannon entropy measures and entropic densities in both position and momentum spaces, highlighting the role of the deformation in the redistribution of quantum information. Finally, our main conclusions are presented in Section~\ref{conclusion}.

\section{The qDFP Model}\label{model}
\begin{figure}[ht]
    \centering
    \includegraphics[width=0.95\linewidth]{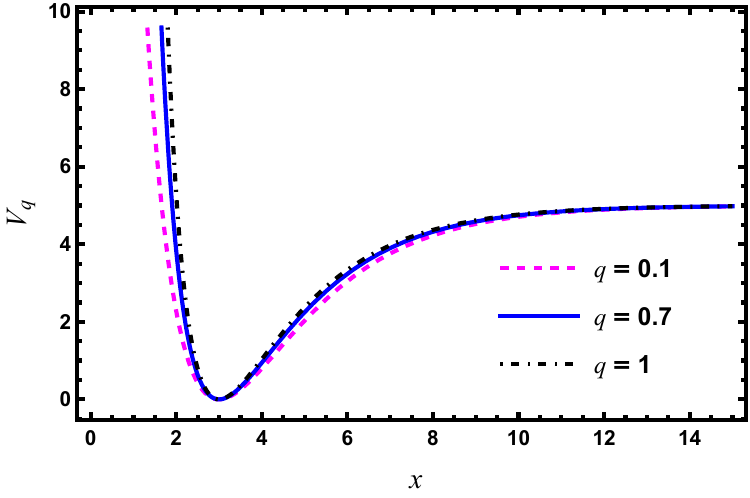}
    \caption{Potential $V_q(x)$ for different values of the deformation parameter $q$. Decreasing $q$ shifts the effective boundary and enhances confinement. $\mathcal{D}_e=5,\, x_e=3,\, \alpha=0.5$. At $q\to 1$, Deng Fan potential is recovered}
    \label{fig:potential}
\end{figure}

The potential 
\begin{eqnarray}
V_q(x)=\mathcal{D}_e\left(\frac{e^{\alpha x}-e^{\alpha x_e}}{e^{\alpha x}-q}\right)^2,
\end{eqnarray}
introduces a deformation parameter $q\in [0.1,1]$ that systematically modifies both the short and long-range behavior of the interaction while recovering the standard Deng-Fan form in the limit $q\to 1$~\cite{deng1957potential}. At short distances, the denominator $e^{\alpha x}-q$ shifts the effective singularity. This produces a steeper repulsive core as $q$ increases, and as such, strengthens confinement. In the asymptotic region $x\to\infty$, the potential admits the expansion
\begin{eqnarray}\label{asymptotic}
V_q(x)\approx \mathcal{D}_e\left(1 + 2\left(q - e^{\alpha x_e}\right)e^{-\alpha x} + \mathcal{O}(e^{-2\alpha x})\right),
\end{eqnarray}
which shows that the long-range tail retains an exponential decay, and the amplitude depends on the combined contribution of $\left(q - e^{\alpha x_e}\right)$. Hence, the quantity $\left(q - e^{\alpha x_e}\right)$ influences the magnitude and nature (attractive or repulsive) of the long-range tail of the potential. Near the equilibrium position $x=x_e$, a Taylor expansion $x=x_e+\Delta x$ gives
\begin{eqnarray}
V_q(x)\approx V_q(x_e)+\frac{1}{2}V_q^{\prime \prime}(x_e)(\Delta x)^2+\cdots,
\end{eqnarray}
where the curvature $V_q^{\prime \prime}(x_e)$ depends explicitly on $q$. This implies that the deformation modifies the local harmonic stiffness of the potential well and hence the vibrational frequency. These results show how $q$ directly contributes to the control of the shape of the potential. As shown in Figure~\ref{fig:potential}, $q<1$, it enhances the short-range repulsion and weakens the long-range attractive tail. It also leads to increased curvature near the equilibrium position and stronger effective confinement. On the other hand, as $q \to 1$, the potential confinement increases, which leads to a narrower and stiffer well around $x_e$. At $q=1$, the curve matches the standard Deng-Fan equilibrium structure.

\subsubsection*{\textit{Molecular admissibility of the $q$DFP Model}}

To ensure that the proposed $q$DFP model remains physically admissible for diatomic molecular systems, we verify that it satisfies the standard conditions for molecular potentials~\cite{wang2012equivalence, jia2012equivalence}. 

First, evaluating the potential at the equilibrium position $x=x_e$ yields
\begin{eqnarray}
V_q(x_e)=D_e\left(\frac{e^{\alpha x_e}-e^{\alpha x_e}}{e^{\alpha x_e}-q}\right)^2=0,
\end{eqnarray}
showing that the minimum occurs at the equilibrium bond length.

Second, the first derivative of the potential vanishes at $x=x_e$, i.e.,
\begin{eqnarray}
\left.\frac{dV_q}{dx}\right|_{x=x_e}=0,
\end{eqnarray}
confirming that this point corresponds to an extremum. Furthermore, the second derivative is positive, that is, the extremum is a true minimum.

Finally, in the asymptotic limit, shown in Eq.~\ref{asymptotic}, the potential approaches
\begin{eqnarray}
\lim_{x\to\infty} V_q(x)=\mathcal{D}_e,
\end{eqnarray}
which corresponds to the correct dissociation energy of the molecule. These results demonstrate that the introduction of the deformation parameter $q$ preserves the essential structural properties required of a molecular potential. 

\section{Exact Solution}\label{exact}
To solve for the exact solution, we adopt the Schr\"odinger equation, 
\begin{eqnarray}\label{schrodinger}
    -\frac{\hbar^2}{2\,m}\frac{d^2 \psi}{dx^2}\,+\,V_q(x) \psi \,=\,\mathcal{E}\psi,
\end{eqnarray}
where $m,\,\hbar$ are in natural units. We introduce the transformation, $\xi = e^{-\alpha x}$, which maps the semi-infinite domain $x\in(0,\infty)$ onto $\xi\in(0,1)$. The radial Schr\"odinger equation reduces to a second-order differential equation with regular singular points at $\xi=0$ and $\xi=1/q$.

Following the standard ansatz~\cite{alici2025brational}
\begin{eqnarray}\label{anstz}
\psi_q(\xi)=\xi^\lambda(1-q\xi)^r f_q(\xi),
\end{eqnarray}
the transformed Eq.~\ref{anstz} is reduced to (See ~\ref{appA})
\begin{eqnarray}\label{eq09}
\xi(1-q \xi)f_q^{\prime \prime}(\xi)+[1+2\lambda-(2\lambda+2r+1)q\xi]f_q^{\prime}(\xi) \\ \nonumber
-\left[q(\lambda+r)^2-\frac{\sigma_1}{q}\right]f_q(\xi)=0,
\end{eqnarray}
which is of Gauss hypergeometric type. $\sigma_1=\beta \,q^2\,+\,\eta\,b^2$, with $b=e^{\alpha x_e}$ and $\eta = 2 m \mathcal{D}_e/\alpha^2\hbar^2$, and $\beta=-2m\mathcal{E}/\alpha^2\hbar^2$. The parameter $\lambda=\sqrt{\beta+\eta}$ governs the asymptotic decay of the wavefunction through its dependence on the energy parameter $\beta$.  $r=1/2\left[1+\sqrt{1+4\delta/q}\,\right],\, 
\delta=\eta/q\big(b(b-2)+q\big)$, contributes to the exponential $r$ and thus governs the behavior of the wavefunction near the finite boundary $\xi=1/q$, induced by the deformation parameter $q$. This separation reflects a key feature of the model that the exponential tail and the near-boundary structure are independently tunable. 

The total wavefunction becomes
\begin{eqnarray}
\psi_{n,q}(\xi)= \mathcal{N}_q\,\xi^\lambda(1-q\xi)^r
{}_2\mathcal{F}_1\big[-n,\;n+2(\lambda+r);\;2\lambda+1;\;q\xi\big],
\end{eqnarray}
where $\xi=e^{-\alpha x}$ and $\mathcal{N}_q$ is the normalization constant.  

Using the quantization condition, the energy eigenvalues are obtained in closed form as:  
\begin{eqnarray}\label{eq:energy}
\mathcal{E}_{n,q}=\mathcal{D}_e-\frac{\hbar^2 \alpha^2}{8m}
\left[
\frac{\eta\left(\frac{b^2}{q^2}-1\right)}{(r+n)}-(r+n)
\right]^2.
\end{eqnarray}
Equation~\ref{eq:energy} shows that the bound-state spectrum is governed by the interplay between the potential depth $\mathcal{D}_e$ and the equilibrium bond length $x_e$, which enter through the terms $\eta\left(\frac{b^2}{q^2}-1\right)$ and $r$, and the quantum number $n$. The deformation parameter $q$ shifts both the level spacing and the dissociation threshold. Also, in the limit $q\to 1$,  the spectrum reduces to the standard Deng-Fan potential~\cite{mesa1998generalized}.

\begin{figure}[ht!]
\centering
\includegraphics[width=0.95\textwidth]{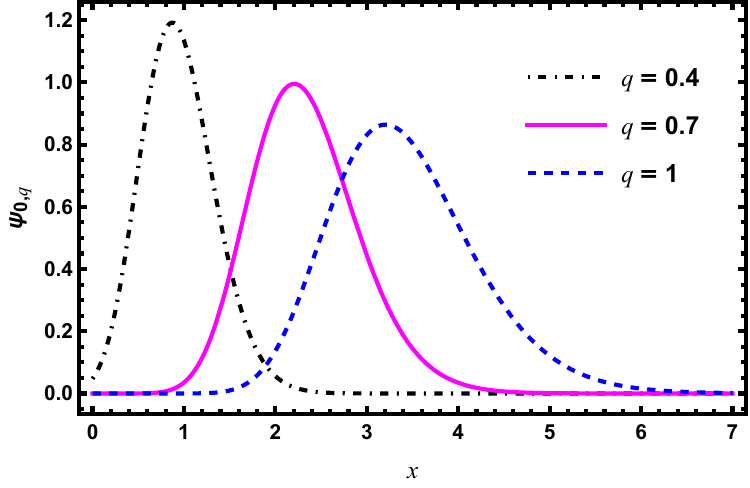}
\caption{Spatial wavefunction of qDFP model in the ground state for  $q\in\{0.4,0.7,1\}$. $\mathcal{D}_e=5,\, \alpha=0.5,\,x_e=3$}
\label{fig:wavefunction_zero}
\end{figure}

\begin{figure}[ht]
\centering
\includegraphics[width=0.95\textwidth]{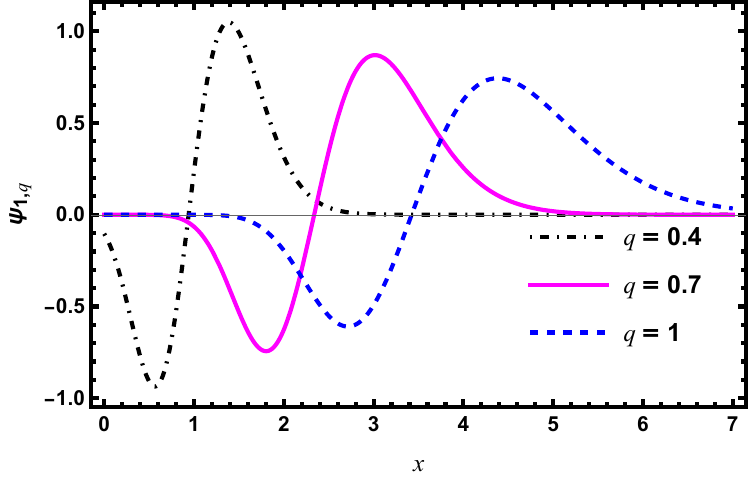}
\caption{Spatial wavefunction in the first excited state for qDFP model for $q\in\{0.4,0.7,1\}\,$. $\mathcal{D}_e=5,\, \alpha=0.5,\,x_e=3$.}
\label{fig:wavefunction_one}
\end{figure}

Fig.~\ref{fig:wavefunction_zero} shows that decreasing $q$ leads to a sharper and more localized ground-state wavefunction around the equilibrium position. The peak height increases while the spatial width decreases, indicating stronger confinement. This behavior is primarily driven by the increase in the effective stiffness of the potential near $x_e$, which compresses the wavefunction rather than shifting it. Furthermore, Fig.~\ref{fig:wavefunction_one},  shows that the first excited state responds more sensitively to $q$ due to its nodal structure. As $q$ decreases, the node shifts toward smaller $x$. Here, both lobes contract, with the outer lobe suppressed more strongly. This asymmetry reflects the reduced support of the wavefunction at larger distances, while the inner region becomes more dominant. Unlike the ground state, where localization is monotonic, the excited state exhibits a redistribution of probability density driven by the combined effect of boundary compression and reduced long-range support.
Fig.~\ref{fig:wavefunction_one} is the spatial wavefunction in the first excited state $(n=1)$. $\psi_{1,q}(\xi)$ exhibits a more sensitive response because of its nodal structure. The introduction of the hypergeometric polynomial in $\psi_{1,q}(\xi)$ introduces a node whose position depends on the balance between $\lambda$ and $r$. As $q<1$, the growth of $r$ shifts the effective boundary at $\xi = 1/q$, therefore, forcing the node closer to the origin. This leads to a visible distortion: both lobes contract, and the outer lobe is more strongly suppressed as a result of the weakened exponential tail described in Section~\ref{model}. Unlike the ground state, where we see that localization is monotonic, the excited state shows competing effects boundary-induced compression that pushes probability inward, and reduced long-range support that suppresses the outer region. The result is a redistribution of probability density toward the inner region, which aligns with the spectral result that higher states are less stable as $q$ decreases.

\section{Quantum Information-Theoretic Measures}\label{quantum}
In this section, we explore Shannon entropy, Fisher information, and Fisher-Shannon complexity measures, to investigate how $q$ distributes information content in conjugate spaces. For a quantum state represented by a normalized wave function $\psi_n(x)$, the probability distribution in configuration space  $\rho_n(x) = |\psi_n(x)|^2$ encodes the likelihood of finding the particle at position $x$. The Shannon entropy in position space is given by
\begin{equation}
S_x^{(n)} =
- \int \rho_n(x)\,\ln\!\big[\rho_n(x)\big]\,dx,
\end{equation}
and serves as a quantitative measure of the spatial spread of the quantum state. On the other hand, the momentum-space representation is obtained via the Fourier transform of the spatial wave function.

The momentum space probability density follows as $\gamma_n(p) = |\psi_n(p)|^2$ and the associated Shannon entropy is 
\begin{equation}
S_p^{(n)} =
- \int \gamma_n(p)\,\ln\!\big[\gamma_n(p)\big]\,dp.
\end{equation}

The position and momentum space Shannon entropy, taken separately decreases (or increases) without bound, and their corresponding probability density becomes more and more (or less) concentrated. The summation of their entropies ($S_x + S_p$), satisfies the Białynicki-Birula and Mycielski (BBM) entropic uncertainty relation as
\begin{equation}\label{BBM}
S_x^{(n)} + S_p^{(n)} \geq D\left(1 + \ln \pi \right),
\end{equation}
establishing a lower bound for the total information entropy~\cite{beckner1975inequalities,bialynicki1975uncertainty}. The boundedness from below of the sum of two entropies means that the total uncertainty in positions and in momentum can not be decreased beyond Eq.~\ref{BBM}. In this analysis, $D=1$.

For a more detailed understanding of how information is distributed across spaces, it is necessary to analyze the corresponding entropy density functions. In the coordinate representation, the local contribution to the Shannon entropy can be expressed as
\begin{align}
\rho^{(n)}_{s}(x) =
\rho_n(x)\,\ln\!\big[\rho_n(x)\big], \\ \nonumber
\rho^{(n)}_{s}(p) =
\gamma_n(p)\,\ln\!\big[\gamma_n(p)\big].
\end{align}

These quantities encode the pointwise contribution of each region in position or momentum space to the total entropy. As such, they offer a more granular description of the information content of the quantum state. By this, we can identify regions that dominate the entropic balance and to better characterize localization and spreading phenomena.

The Fisher information in position and momentum space
\begin{eqnarray}
   I_x&=& \int{\frac{|\nabla \rho(x)|^2}{\rho(x)}}=4\int{|\psi^{\prime}(x)|^2}\,dx, \\ \nonumber
   I_p&=&  \int{\frac{|\nabla \gamma(p)|^2}{\gamma(p)}}=4\int{|\psi^{\prime}(p)|^2}\,dp
\end{eqnarray}
The Cramér–Rao uncertainty relation for the Fisher information\cite{romera2005fisher}
\begin{equation}
    I_x\,I_p \ge 36
\end{equation}

Complexity measures (CM) are statistical tools that combine local and global variables from information-theoretic measures. CM acts as bridge between complete order (minimum entropy) and total disorder (maximum entropy) to provide structural insight of the quantum system. For Fisher and Shannon complexity, it takes the form~\cite{lopez2011statistical}
\begin{eqnarray}
 C_{\mathcal{FS}}^{i}&=& J\cdot I ,\qquad i\in\{x,p\}, \\ \nonumber
 J_i&=&  \frac{1}{2 \pi e}\,e^{\frac{2}{D} S_i},  
\end{eqnarray}
where $J$ is Shannon entropy power~\cite{dembo1991information}, $D=1$, and their product, $J\cdot I \ge1$ 

\begin{figure}[ht!]
\begin{center}
\begin{tabular}{ccc}
\includegraphics[height=6cm]{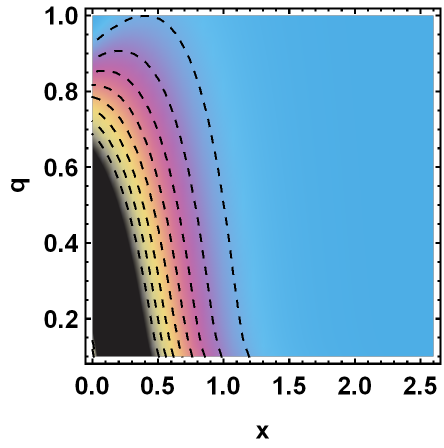} 
\includegraphics[height=6cm]{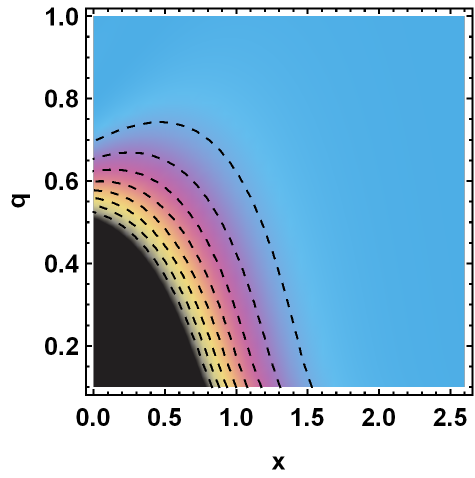}
\includegraphics[height=6cm]{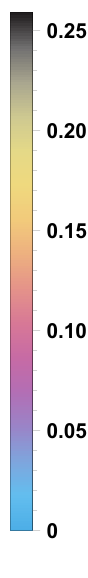}
\end{tabular}
\end{center}
\caption{Entropic information density for $\mathcal{D}_e=5,\, \alpha=0.5,\,x_e=3$ and $m=10^{-6}$. 
The plot on the left refers to the ground state with energy $n=0$. The plot on the right is for the first excited state $n=1$ (Here we consider $\hbar=1$).
\label{fig04}}
\end{figure}

Figure~\ref{fig04} displays the spatial distribution of the entropic information density $\rho^{(n)}_{s}(x)$ for the $q$DFP model, for ground state ($n=0$, left panel) and the first excited state ($n=1$, right panel). In the ground state, the entropic density concentrates sharply near the inner boundary of the potential and shifts further inward as $q$ decreases. This behavior follows directly from the steepening of the repulsive core: stronger deformation compresses the wavefunction, raising the local probability density and therefore the local entropy contribution in the inner region. The ground state response is monotonic; decreasing $q$ unambiguously transfers information content from the asymptotic tail toward the equilibrium region. On the other hand, nodal structure $\psi_{1,q}$ splits the entropic density into two spatially separated regions, and these two regions respond to $q$ unequally. As $q$ decreases, the outer lobe loses support faster than the inner lobe gains it, so the net effect is a redistribution rather than a simple compression. This asymmetry reflects the reduced long-range support of the wavefunction discussed in Section~\ref{exact}, and it explains why excited states are informationally more sensitive to deformation than the ground state. Figure~\ref{fig04} establishes the fact that $q$ does not act uniformly across quantum states: it compresses ground state information monotonically, but reshapes excited state information through a competition between boundary compression and reduced asymptotic support.
\begin{table}[htbp]
\centering
\caption{Numerical Shannon entropy for for $\mathcal{D}_e=5,\, \alpha=0.5,\,x_e=3$ and $m=10^{-6}$ (Here we consider $\hbar=1$).}
\label{table:one}
\begin{tabular}{lccccc}
\hline
       $n$ & $q$  & $S_{x}$ & $S_{p}$ & $S_{x}+S_{p}$    \\
\hline
        $0$ & $0.3$  & $-0.42404$ & $3.89058$ & $3.46654$\\
            & $0.5$  & $-0.34183$ & $3.79122$ & $3.44938$\\
            & $0.8$  & $-0.05544$ & $3.38737$ & $3.33193$\\
            & $1.0$  & $0.24720$ & $2.67318$ & $2.92038$\\
\hline
        $1$ & $0.3$  & $-0.19213$ & $3.62274$ & $3.43061$\\
            & $0.5$  & $0.04672$ & $3.29821$ & $3.34493$\\
            & $0.8$  & $0.53755$ & $2.24704$ & $2.78459$\\
            & $1.0$  & $0.65044$ & $2.16158$ & $2.81202$\\
\hline
        $2$ & $0.3$  & $-0.09595$ & $3.49328$ & $3.39733$\\
            & $0.5$  & $0.33780$ & $2.81759$ & $3.15539$\\
            & $0.8$  & $0.70983$ & $2.16635$ & $2.87618$\\
            & $1.0$  & $0.80447$ & $2.06991$ & $2.87438$\\
\hline
\end{tabular}
\end{table}

Table~\ref{table:one}  presents the Shannon entropies in position space $S_x$ and momentum space $S_p$, and their sum for quantum states $n=\{0,1,2\}$ across the deformation parameters $ q\in \{0.3,0.5,0.8, 0.1\}$. The BBM lower bound for $D=1$ gives $S_x +S_p \ge 1+\ln \pi \approx 2.144$,  and all computed values satisfy this bound. These results confirm the physical admissibility of the obtained states. The individual entropies reveal a strict complementarity governed by $q$.  As $q\to1$, it increases towards the standard Deng-Fan limit. As seen in Table~\ref{table:one}, a clear trend emerges as $q$ increases: the position-space entropy $S_x$ increases, indicating a progressive delocalization of the wavefunction in position space. Also, The negative values of $S_x$ at small $q$ arise because strong spatial confinement drives the probability density $\rho(x)= |\psi(x)|^2$, to values much greater than unity in the localized region. This is so that $\ln \rho > 0$ and the integrand $- \rho \ln \rho <0$, which makes $S_x$ negative. This is a well-documented feature of continuous probability distributions under extreme localization and has been reported in comparable confinement studies~\cite{aquino2013shannon}. Conversely, the momentum-space entropy $S_p$ decreases, reflecting a reduction in the uncertainty in momentum space. A more physically significant result lies in the total entropy $S_x +S_p$ and its significance to information content. The BBM bound defines the minimum total uncertainty achievable under standard Fourier duality. Any excess, $\Delta (q) =(S_x +S_p)-(1+\ln\pi)$, above this BBM bound represents genuine irreducible ignorance about complementary observables beyond the quantum mechanical minimum ~\cite{bialynicki1975uncertainty}. From our analysis, for $n=0: \Delta (0.3)=0.1322; \Delta(1.0)=0.776$, we see that at $q\to1$, $\Delta(q)$ decreases which increases total uncertainty about the joint position-momentum description. This means that the system knows more about the location of the particle. Decreasing $q$ pushes the quantum state further from the minimum-uncertainty configuration,  and redistributes total information. Furthermore, the quantum number $n$ amplifies this effect. Higher excited states carry broader spatial profiles and therefore larger $S_x$ but their $S_p$ decreases correspondingly. The entropic excess $\Delta$ remains substantial accross all $n$, which confirms that the deformation reshapes information content at every level of excitation. 

An important boundary condition governs the validity of these results. The mass parameter $m=10^{-6}$ is adopted for this analysis.  At $m\to 1$, the BBM inequality is no longer valid for all values of $q$ and $\Delta(q)$ becomes negative. This indicates that the underlying assumptions of standard Fourier duality between position and momentum spaces break down under the combined effect of the deformation and the mass scaling. Also, the probability densities in position and momentum space cease to maintain the mathematical relationship that the BBM bound requires. In this sense, the parameter $m$ plays a crucial role in controlling the validity of the information-theoretic description, with small $m$ guaranteeing physically meaningful and well-behaved quantum states.

\begin{figure}[ht!]
\begin{center}
\begin{tabular}{ccc}
\includegraphics[height=6cm]{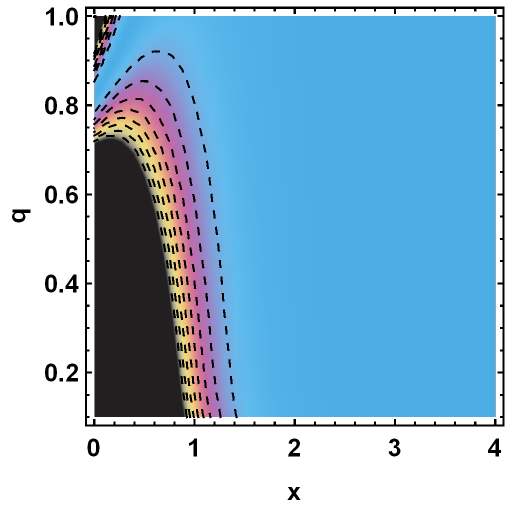} 
\includegraphics[height=6cm]{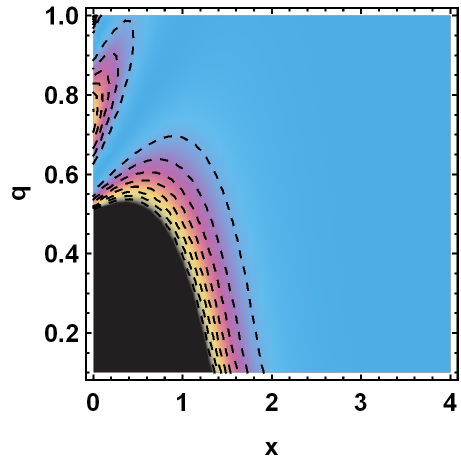}
\includegraphics[height=6cm]{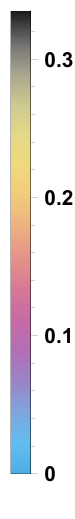}
\end{tabular}
\end{center}
\caption{ Fisher information density for $\mathcal{D}_e=5,\, \alpha=0.5,\,x_e=3$ and $m=10^{-6}$. 
The plot on the left refers to the ground state with energy $n=0$. The plot on the right is for the first excited state $n=1$ (Here we consider $\hbar=1$).
\label{fig11}}
\end{figure}

Figure~\ref{fig11} shows the Fisher information density in position space for the $q$DFP model, for both the ground state ($n=0$, left panel) and the first excited state ($n=1$, right panel). Unlike the Shannon entropy density, which measures the spread of the probability distribution, the Fisher information is highly sensitive to local variations and gradients of the wavefunction. As a result, regions where the wavefunction exhibits rapid spatial changes contribute most significantly to the Fisher information. For the ground state, the Fisher information density is strongly concentrated near the inner region of the potential, where the wavefunction varies most rapidly due to the steep repulsive core. As the deformation parameter $q$ decreases, this concentration becomes more pronounced and shifts toward smaller values of $x$, indicating an increase in localization and sharper spatial features. In the first excited state, the presence of a node introduces additional structure, leading to multiple regions of high Fisher information. In particular, the vicinity of the node and the inner boundary both contribute significantly, and reflects the enhanced sensitivity of Fisher information to nodal and boundary-induced variations. 

\begin{table}[htbp]
\centering
\caption{Numerical Fisher information for $\mathcal{D}_e=5$,\, $\alpha=0.5$,\,$x_e=3$ and $m=10^{-6}$ (Here we consider $\hbar=1$).}
\label{tab12}
\begin{tabular}{lccccc}
\hline
       $n$ & $q$  & $I_{x}$ & $I_{p}$ & $I_{x}I_{p}$    \\
\hline
        $0$ & $0.3$  & $16.8411$ & $8.3138$ & $140.0150$\\
            & $0.5$  & $13.5868$ & $6.7221$ & $91.3327$\\
            & $0.8$  & $10.3629$ & $4.6806$ & $48.5055$\\
            & $1.0$  & $24.7026$ & $11.9622$ & $295.4980$\\
\hline
        $1$ & $0.3$  & $9.9902$ & $12.4093$ & $123.9721$\\
            & $0.5$  & $6.9133$ & $6.1405$ & $42.4518$\\
            & $0.8$  & $8.7569$ & $10.9070$ & $95.5118$\\
            & $1.0$  & $23.9789$ & $29.1911$ & $699.9710$\\
\hline
        $2$ & $0.3$  & $7.5797$ & $9.4381$ & $71.5386$\\
            & $0.5$  & $8.3658$ & $10.9580$ & $91.6740$\\
            & $0.8$  & $12.6398$ & $15.7566$ & $199.1606$\\
            & $1.0$  & $29.8448$ & $36.0482$ & $1075.8554$\\
\hline
\end{tabular}
\end{table}

Table~\ref{tab12} presents the numerical results for the Fisher information in position space ($I_x$), momentum space ($I_p$), and their product for different quantum states $n=\{0,1,2\}$ and deformation parameters $q$. A clear dependence on the deformation parameter is observed, a reflection on the sensitivity of Fisher information to local variations of the wavefunction. For small values of $q$, the relatively large values of $I_x$ indicate strong spatial localization. This is associated with steep gradients of the wavefunction near the inner region of the potential. This is accompanied by moderate values of $I_p$, consistent with a broader distribution in momentum space. As $q$ increases toward the Deng--Fan limit ($q\to 1$), both $I_x$ and $I_p$ tend to increase, especially for excited states. This indicates that the wavefunction develops sharper features in both spatial representations. The product $I_x I_p$ satisfies the Cramér--Rao inequality for all cases considered, consistent with the results within the information-theoretic framework. Notably, the product grows substantially with increasing $q$ and quantum number $n$, which reveals that higher excited states and weaker deformation lead to enhanced overall information content, characterized by stronger fluctuations in both position and momentum spaces. This behavior highlights the dual role of the deformation parameter: while smaller $q$ values promote spatial localization, larger $q$ values induce more complex oscillatory structures, particularly in excited states, which are captured by the Fisher information. 

\begin{table}[htbp]
\centering
\caption{Numerical complexity measures for $\mathcal{D}_e=5$,\, $\alpha=0.5$,\,$x_e=3$ and $m=10^{-6}$ (Here we consider $\hbar=1$).}
\label{tab13}
\begin{tabular}{lccccc}
\hline
       $n$ & $q$  & $ C_{\mathcal{FS}}^{x}$ & $ C_{\mathcal{FS}}^{p}$     \\
\hline
        $0$ & $0.3$  & $1.14781$ & $3169.16$ \\
            & $0.5$  & $1.09151$ & $2100.56$ \\
            & $0.8$  & $1.08688$ & $373.56$ \\
            & $1.0$  & $6.44593$ & $399.50$ \\
\hline
        $1$ & $0.3$  & $1.08270$ & $2768.47$ \\
            & $0.5$  & $1.03333$ & $715.83$ \\
            & $0.8$  & $4.08397$ & $155.34$ \\
            & $1.0$  & $14.0156$ & $350.42$ \\
\hline
        $2$ & $0.3$  & $1.12706$ & $1625.31$ \\
            & $0.5$  & $1.05277$ & $131.87$ \\
            & $0.8$  & $8.31972$ & $190.96$ \\
            & $1.0$  & $23.7382$ & $360.25$ \\
\hline
\end{tabular}
\end{table}

Table~\ref{tab13} presents the Fisher--Shannon complexity measures in position space ($C_{\mathcal{FS}}^{x}$) and momentum space ($C_{\mathcal{FS}}^{p}$) for different quantum states and deformation parameters $q$. These quantities combine global (Shannon entropy) and local (Fisher information) features to provide a more complete characterization of the structural complexity of the quantum states. In position space, $C_{\mathcal{FS}}^{x}$ remains close to its lower bound for small values of $q$, indicating that although the states are highly localized, their internal structure is relatively simple. However, as $q$ increases, particularly toward $q=1$, $C_{\mathcal{FS}}^{x}$ grows significantly, especially for excited states. This reflects the emergence of more intricate spatial patterns associated with increased oscillatory behavior and reduced confinement. In momentum space, the behavior is marked differently: $C_{\mathcal{FS}}^{p}$ assumes very large values for small $q$, which signals a highly complex distribution characterized by strong delocalization combined with significant gradient variations. As $q$ increases, $C_{\mathcal{FS}}^{p}$ decreases substantially, indicating a simplification of the momentum-space structure. This complementary behavior between position and momentum spaces highlights the trade-off between localization and structural complexity. Furthermore, all values satisfy the inequality $C_{\mathcal{FS}}^{i} \geq 1$, consistent with the theoretical bounds. Our results demonstrate that $q$ plays a crucial role in controlling localization and also in tuning the intrinsic complexity of the quantum states, with excited states exhibiting a stronger sensitivity to these effects.

\section{Conclusion}\label{conclusion}
In this work, we present an exactly solvable $q$DFP that extends the standard DFP model by introducing a single $ q$-deformation parameter that controls both spectral structure and spatial localization. The analytical solution reveals that the deformation alters the effective quantum number and energy scale in a coupled manner, which produces non-uniform spectral shifts and compression toward the dissociation limit. The modification of the repulsive core and asymptotic tail leads to an improved localization and reduced support for higher excited states. The $qDFP$ model preserves molecular admissibility and recovers the standard Deng-Fan limit as $q \to 1$, which provides a consistent framework for studying deviations from standard intermolecular interactions.

From the perspective of quantum information, the $q$DFP model provides a clear and quantitative link between deformation and information flow in quantum states. The Shannon entropy analysis shows that $q$ governs the redistribution of information between position and momentum spaces in a controlled manner. Decreasing $q$ enhances spatial localization, leading to lower and even negative values of  $S_x$, while simultaneously increasing $S_p$, which reflects stronger momentum delocalization. Negative values arise because strong confinement drives $\rho(x) >>1$; a physical consequences of confinement. This complementary behavior is fully consistent with the BBM entropic uncertainty relation, whose validity confirms the physical reliability of the obtained states in the regime $m \ll 1$, where standard Fourier duality between position and momentum spaces is preserved. Critically, stronger deformation pushes the quantum state further from the minimum-uncertainty configuration: the entropic excess $\Delta (q) =(S_x +S_p)-(1+\ln\pi)$ grows as $q$ decreases. This means the system gains positional knowledge at the cost of increased total informational uncertainty about complementary observables. The entropy density analysis further shows that deformation suppresses asymptotic contributions and concentrates local information content near the equilibrium position, with excited states exhibiting stronger sensitivity through their nodal structures.

Complementing the Shannon entropy analysis, the Fisher information and Fisher-Shannon complexity measures characterize the local and structural properties of the quantum states. Smaller $q$ increases $I_x$ and redistributes information toward momentum space, while the Cramér-Rao inequality is satisfied in all cases. The Fisher-Shannon complexity reveals a clear trade-off: for any $n$ state, strongly deformed regimes exhibit low structural complexity in position space and high in momentum space. Therefore, the Shannon entropy, Fisher information, and their complexity results establish a unified picture: the $q$-deformation redistributes information, reshaping quantum states in ways that purely spectral analysis cannot capture.

\section*{Acknowledgment}
PA and PK acknowledge the funding support from the NSRF via the Research and Innovation Acceleration Agency for Competitiveness and Area 4/5 Development (RCAD) (Program Management Unit for Frontier Brainpower and Future Industries) [grant number B39G690076]. COE and NA  would also like to acknowledge the Universiti Malaysia Perlis (Funding No. 9004-00100 Special Research Grant-International Postdoctoral) for funding this project. 
\section*{Conflict of interest} The authors declare that they have no known competing financial interests or personal relationships that could have appeared to influence the work reported in this paper
\appendix
\section{Exact Solution of DFP Model}\label{appA}
From Equation~\ref{schrodinger}
\begin{eqnarray}\label{app1}
    \frac{d^2\psi(x)}{dx^2}+\frac{2m}{\hbar^2}\left[\mathcal{E}-V_q\right]\psi(x)=0
\end{eqnarray}
We perform coordinate transformation. Let $\xi =e^{-\alpha x}$ and with algebraic arranagement, we have

\begin{eqnarray}
    \xi^2 \frac{d^2\psi(\xi)}{d\xi^2}+\xi \frac{d\psi(\xi)}{d\xi}+\left[\beta-\frac{\eta\,(1-b \xi)^2}{(1-q\xi)^2}\right]\psi(\xi)=0,
\end{eqnarray}
where $b=e^{\alpha x_e},\,\eta=2 m\mathcal{D}_e/\alpha^2\hbar^2$ and $\beta=-2m\mathcal{E}/\alpha^2\hbar^2$. factorizing $\xi(1-q\,\xi)$, we obtain
\begin{eqnarray}
    \xi(1-q\xi)\frac{d^2\psi(\xi)}{d\xi^2}+(1-q\xi)\frac{d\psi(\xi)}{d\xi}+\frac{1}{\xi(1-q\xi)}\left[-\beta(1-q\xi)^2-\eta(1-b\xi)^2\right]\psi(\xi)=0
\end{eqnarray}
\begin{eqnarray}\label{eq21}
     \xi(1-q\xi)\frac{d^2\psi(\xi)}{d\xi^2}+(1-q\xi)\frac{d\psi(\xi)}{d\xi}+\frac{1}{\xi(1-q\xi)}\left[-\sigma_1\xi^2+\sigma_2\xi-\sigma_3\right]\psi(\xi)=0,
\end{eqnarray}
where $\sigma_1=(\beta q^2+\eta b ^2),\, \sigma_2=2\beta+2\eta b,\, \sigma_3=\beta+\eta$. We assume the wavefunction for the $q$DFP model
\begin{eqnarray}
    \psi_q(\xi)=\xi^\lambda(1-q\xi)^r f_q(\xi)
\end{eqnarray}

Substituting into Eq.~\ref{eq21} and performing algebraic manipulation, we  get
\begin{eqnarray}\label{ap23}
\xi(1-q \xi)f_q^{\prime \prime}+\left[1+2\lambda-(2\lambda+2r+1)q\xi\right]f_q^{\prime}- \\ \nonumber
\left[2q\lambda r+q\lambda^2+r^2q-\frac{\sigma_1}{q}\,+\,\frac{\lambda^2-\sigma_3}{\xi}+\frac{ q r(r-1)-\sigma_1/q+\sigma_2-\sigma_3 q}{(1-q\xi)}\right]f_q=0,
\end{eqnarray}
with singularity at $\xi=0$ and $\xi=1/q$. 

Solving for $\lambda$ and $r$, we get
\begin{eqnarray}
  \lambda=\sqrt{-\frac{2m\mathcal{E}}{\alpha^2\hbar^2}+\frac{2m\mathcal{D}_e}{\alpha^2\hbar^2}}, \\ \nonumber
  r=\frac{1}{2}\left[1+\sqrt{1+\frac{4 \delta}{q}}]\right],\quad \delta=\frac{\eta}{q}\left[b(b-2)+q\right]
\end{eqnarray}
Equation~\ref{ap23} becomes
\begin{eqnarray}
    \xi(1-q \xi)f_q^{\prime \prime}(\xi)+[1+2\lambda-(2\lambda+2r+1)q\xi]f_q^{\prime}(\xi) \\ \nonumber
-\left[q(\lambda+r)^2-\frac{\sigma_1}{q}\right]f_q(\xi)=0,
\end{eqnarray}
which is a representative of the standard Gauss hypergeometric equation, which takes the form.
\begin{eqnarray}
z(1-z)f^{\prime \prime}(z)+[c-(a+b+1)z]f^{\prime}(z)-ab\,f(z)=0.
\end{eqnarray}
Introducing the transformation $z=q\xi$. By direct comparison with Eq.~\ref{eq09}, the parameters are identified as
\begin{eqnarray}
c=1+2\lambda,\quad a+b=2(\lambda+r),\quad ab=(\lambda+r)^2-\frac{\sigma_1}{q^2}.
\end{eqnarray}
Solving $a$ and $b$, we obtain
\begin{eqnarray}
a=(\lambda+r)-\frac{\sqrt{\sigma_1}}{q}, \qquad 
b=(\lambda+r)+\frac{\sqrt{\sigma_1}}{q}.
\end{eqnarray}

The regular solution at $z=0$ is therefore expressed as
\begin{eqnarray}\label{A4}
f_q(\xi)={}_2F_1\!\left((\lambda+r)-\frac{\sqrt{\sigma_1}}{q},\;
(\lambda+r)+\frac{\sqrt{\sigma_1}}{q};\;
1+2\lambda;\;q\xi\right).
\end{eqnarray}

The requirement of normalizability imposes the condition that the hypergeometric series terminates, which occurs when $a=-n$, with $n=0,1,2,\dots$. This leads to the quantization condition  
\begin{eqnarray}\label{app13}
\frac{\sqrt{\sigma_1}}{q}=\lambda+r+n.
\end{eqnarray}

From Eq.~\ref{A4}, the wavefunction then reduces to the polynomial form  
\begin{eqnarray}
f_q(\xi)={}_2F_1\big(-n,\;n+2(\lambda+r);\;2\lambda+1;\;q\xi\big),
\end{eqnarray}
and the total wavefunction becomes  
\begin{eqnarray}
\psi_{n,q}(\xi)= \mathcal{N}_q\,\xi^\lambda(1-q\xi)^r
{}_2F_1\big[-n,\;n+2(\lambda+r);\;2\lambda+1;\;q\xi\big],
\end{eqnarray}
where $\xi=e^{-\alpha x}$ and $\mathcal{N}$ is the normalization constant defined as:
\begin{eqnarray}
    \mathcal{N}_q = \left[\int|\psi_{n,q}(x)|\,dx\right]^{-1/2}
\end{eqnarray}
The normalization constant $ \mathcal{N}_q$ is evaluated numerically for all quantum states considered.


\begin{thebibliography}{10}
\expandafter\ifx\csname url\endcsname\relax
  \def\url#1{\texttt{#1}}\fi
\expandafter\ifx\csname urlprefix\endcsname\relax\def\urlprefix{URL }\fi
\expandafter\ifx\csname href\endcsname\relax
  \def\href#1#2{#2} \def\path#1{#1}\fi

\bibitem{hajigeorgiou2010extended}
P.~G. Hajigeorgiou, An extended lennard-jones potential energy function for diatomic molecules: Application to ground electronic states, Journal of Molecular Spectroscopy 263~(1) (2010) 101–110.
\newblock \href {https://doi.org/https://doi.org/10.1016/j.jms.2010.07.003} {\path{doi:https://doi.org/10.1016/j.jms.2010.07.003}}.

\bibitem{deng1957potential}
Z.~H. Deng, Y.~Fan, A potential function of diatomic molecules, Journal of Shandong University 1~(11) (1957).

\bibitem{mesa1998generalized}
A.~D.~S. Mesa, C.~Quesne, Y.~F. Smirnov, Generalized morse potential: Symmetry and satellite potentials, Journal of Physics A: Mathematical and General 31~(1) (1998) 321–335.
\newblock \href {https://doi.org/https://doi.org/10.1088/0305-4470/31/1/028} {\path{doi:https://doi.org/10.1088/0305-4470/31/1/028}}.

\bibitem{morse1929diatomic}
P.~M. Morse, Diatomic molecules according to the wave mechanics. ii. vibrational levels, Physical Review 34~(1) (1929) 57–64.
\newblock \href {https://doi.org/https://doi.org/10.1103/physrev.34.57} {\path{doi:https://doi.org/10.1103/physrev.34.57}}.

\bibitem{rong2003comparison}
Z.~Rong, H.~G. Kjaergaard, M.~L. Sage, Comparison of the morse and deng-fan potentials for x-h bonds in small molecules, Molecular Physics 101~(14) (2003) 2285–2294.
\newblock \href {https://doi.org/https://doi.org/10.1080/0026897031000137706} {\path{doi:https://doi.org/10.1080/0026897031000137706}}.

\bibitem{muser2022interatomic}
M.~H. Müser, S.~V. Sukhomlinov, L.~Pastewka, Interatomic potentials: achievements and challenges, Advances in Physics: X 8~(1) (Nov 2022).
\newblock \href {https://doi.org/https://doi.org/10.1080/23746149.2022.2093129} {\path{doi:https://doi.org/10.1080/23746149.2022.2093129}}.

\bibitem{dunham1932energy}
J.~L. Dunham, The energy levels of a rotating vibrator, Physical Review 41~(6) (1932) 721–731.
\newblock \href {https://doi.org/https://doi.org/10.1103/physrev.41.721} {\path{doi:https://doi.org/10.1103/physrev.41.721}}.

\bibitem{vogt2018accuracy}
E.~Vogt, D.~S. Sage, H.~G. Kjaergaard, Accuracy of xh-stretching intensities with the deng–fan potential, Molecular physics 117~(13) (2018) 1629–1639.
\newblock \href {https://doi.org/https://doi.org/10.1080/00268976.2018.1521529} {\path{doi:https://doi.org/10.1080/00268976.2018.1521529}}.

\bibitem{wang2012equivalence}
P.-Q. Wang, L.-H. Zhang, C.-S. Jia, J.-Y. Liu, Equivalence of the three empirical potential energy models for diatomic molecules, Journal of Molecular Spectroscopy 274 (2012) 5–8.
\newblock \href {https://doi.org/https://doi.org/10.1016/j.jms.2012.03.005} {\path{doi:https://doi.org/10.1016/j.jms.2012.03.005}}.

\bibitem{alici2025brational}
H.~Alıcı, S.~Ulusoy, On vibrational levels of the deng-fan molecular potential, Physica Scripta 100~(4) (2025) 045239.
\newblock \href {https://doi.org/https://doi.org/10.1088/1402-4896/adc15f} {\path{doi:https://doi.org/10.1088/1402-4896/adc15f}}.

\bibitem{oyewumi2013thermodynamic}
K.~Oyewumi, B.~Falaye, C.~Onate, O.~Oluwadare, W.~Yahya, Thermodynamic properties and the approximate solutions of the schrödinger equation with the shifted deng–fan potential model, Molecular Physics 112~(1) (2013) 127–141.
\newblock \href {https://doi.org/https://doi.org/10.1080/00268976.2013.804960} {\path{doi:https://doi.org/10.1080/00268976.2013.804960}}.

\bibitem{dong2008arbitary}
S.-H. Dong, X.-Y. Gu, Arbitrary l state solutions of the schrödinger equation with the deng-fan molecular potential, Journal of Physics Conference Series 96~(012109) (Feb 2008).
\newblock \href {https://doi.org/https://doi.org/10.1088/1742-6596/96/1/012109} {\path{doi:https://doi.org/10.1088/1742-6596/96/1/012109}}.

\bibitem{greene1976variation}
R.~L. Greene, C.~Aldrich, Variational wave functions for a screened coulomb potential, Physical Review A 14~(6) (1976) 2363–2366.
\newblock \href {https://doi.org/https://doi.org/10.1103/physreva.14.2363} {\path{doi:https://doi.org/10.1103/physreva.14.2363}}.

\bibitem{oyewumi2012bound}
K.~J. Oyewumi, O.~J. Oluwadare, K.~D. Sen, O.~A. Babalola, Bound state solutions of the deng–fan molecular potential with the pekeris-type approximation using the nikiforov–uvarov (n–u) method, Journal of Mathematical Chemistry 51~(3) (2012) 976–991.
\newblock \href {https://doi.org/https://doi.org/10.1007/s10910-012-0123-6} {\path{doi:https://doi.org/10.1007/s10910-012-0123-6}}.

\bibitem{zhang2011approximate}
L.~Zhang, X.~Li, C.~Jia, Approximate solutions of the schrödinger equation with the generalized morse potential model including the centrifugal term, International Journal of Quantum Chemistry 111~(9) (2011) 1870–1878.
\newblock \href {https://doi.org/https://doi.org/10.1002/qua.22477} {\path{doi:https://doi.org/10.1002/qua.22477}}.

\bibitem{abu-shady2022general}
M.~Abu-Shady, E.~M. Khokha, T.~A. Abdel-Karim, The generalized fractional nu method for the diatomic molecules in the deng–fan model, The European Physical Journal D 76~(9) (Sep 2022).
\newblock \href {https://doi.org/https://doi.org/10.1140/epjd/s10053-022-00480-w} {\path{doi:https://doi.org/10.1140/epjd/s10053-022-00480-w}}.

\bibitem{hassanabadi2012deng}
H.~Hassanabadi, B.~Yazarloo, S.~Zarrinkamar, H.~Rahimov, Deng-fan potential for relativistic spinless particles — an ansatz solution, Communications in Theoretical Physics 57~(3) (2012) 339–342.
\newblock \href {https://doi.org/https://doi.org/10.1088/0253-6102/57/3/02} {\path{doi:https://doi.org/10.1088/0253-6102/57/3/02}}.

\bibitem{ikot2013solution}
A.~N. Ikot, O.~A. Awoga, Solutions of dirac equation with generalized rotating deng-fan potential, Arabian Journal for Science and Engineering 39~(1) (2013) 467–474.
\newblock \href {https://doi.org/https://doi.org/10.1007/s13369-013-0829-1} {\path{doi:https://doi.org/10.1007/s13369-013-0829-1}}.

\bibitem{halder2025information}
A.~Halder, A.~K. Roy, D.~Nath, Information theoretic measures within schr$\backslash$" odinger-dunkl framework in spherical coordinates, arXiv preprint arXiv:2506.17447 (2025).

\bibitem{onate2018effect}
C.~Onate, A.~Ikot, M.~Onyeaju, O.~Ebomwonyi, J.~Idiodi, Effect of dissociation energy on shannon and rényi entropies, Karbala International Journal of Modern Science 4~(1) (2018) 134–142.
\newblock \href {https://doi.org/https://doi.org/10.1016/j.kijoms.2017.12.004} {\path{doi:https://doi.org/10.1016/j.kijoms.2017.12.004}}.

\bibitem{nyeo2001phase}
S.-L. Nyeo, I.-C. Yang, Phase transition of a quasi-one-dimensional system, Physical Review E 63~(4) (Mar 2001).
\newblock \href {https://doi.org/https://doi.org/10.1103/physreve.63.046109} {\path{doi:https://doi.org/10.1103/physreve.63.046109}}.

\bibitem{hamzavi2012equivalence}
M.~Hamzavi, S.~M. Ikhdair, K.~Thylwe, Equivalence of the empirical shifted deng–fan oscillator potential for diatomic molecules, Journal of Mathematical Chemistry 51~(1) (2012) 227–238.
\newblock \href {https://doi.org/https://doi.org/10.1007/s10910-012-0075-x} {\path{doi:https://doi.org/10.1007/s10910-012-0075-x}}.

\bibitem{edetikot2022superstatistics}
C.~Edet, A.~Ikot, Superstatistics of diatomic molecules with the shifted deng-fan potential model, Biointerface Research in Applied Chemistry 12~(3) (2022) 4126–4139.
\newblock \href {https://doi.org/https://doi.org/10.33263/briac123.41264139} {\path{doi:https://doi.org/10.33263/briac123.41264139}}.

\bibitem{ghanbari2025theoretical}
A.~Ghanbari, Theoretical calculations of thermal functions of diatomic molecules using shifted deng-fan potential, Computational and Theoretical Chemistry 1248 (2025) 115186.
\newblock \href {https://doi.org/https://doi.org/10.1016/j.comptc.2025.115186} {\path{doi:https://doi.org/10.1016/j.comptc.2025.115186}}.

\bibitem{lumb2016rovibrational}
S.~Lumb, S.~Lumb, V.~Prasad, Rovibrational spectra of bounded diatomic molecules, International Journal of Quantum Chemistry 117~(6) (2016) e25339.
\newblock \href {https://doi.org/https://doi.org/10.1002/qua.25339} {\path{doi:https://doi.org/10.1002/qua.25339}}.

\bibitem{oluwadare2018energy}
O.~J. Oluwadare, K.~J. Oyewumi, Energy spectra and the expectation values of diatomic molecules confined by the shifted deng-fan potential, The European Physical Journal Plus 133~(10) (Oct 2018).
\newblock \href {https://doi.org/https://doi.org/10.1140/epjp/i2018-12210-0} {\path{doi:https://doi.org/10.1140/epjp/i2018-12210-0}}.

\bibitem{angelova2004revisiting}
M.~N. Angelova, V.~K. Dobrev, A.~Frank, Revisiting the quantum group symmetry of diatomic molecules, The European Physical Journal D 31 (2004) 27–37.
\newblock \href {https://doi.org/10.1140/epjd/e2004-00111-6} {\path{doi:10.1140/epjd/e2004-00111-6}}.

\bibitem{hassanabadi2017deformed}
H.~Hassanabadi, W.~S. Chung, S.~Zare, S.~B. Bhardwaj, Q-deformed morse and oscillator potential, Advances in High Energy Physics 2017 (2017).
\newblock \href {https://doi.org/https://doi.org/10.1155/2017/1730834} {\path{doi:https://doi.org/10.1155/2017/1730834}}.

\bibitem{boumali2018statistical}
A.~Boumali, The statistical properties of q-deformed morse potential for some diatomic molecules via euler–maclaurin method in one dimension, Journal of Mathematical Chemistry 56~(6) (2018) 1656–1666.
\newblock \href {https://doi.org/10.1007/s10910-018-0879-4} {\path{doi:10.1007/s10910-018-0879-4}}.

\bibitem{yan1990deformed}
H.~Yan, q-deformed oscillator algebra as a quantum group, Journal of Physics A: Mathematical and General 23 (1990) L1155--L1160.
\newblock \href {https://doi.org/10.1088/0305-4470/23/22/001} {\path{doi:10.1088/0305-4470/23/22/001}}.

\bibitem{schmidt2006deformed}
A.~Schmidt, H.~Wachter, q-deformed quantum lie algebras, Journal of Geometry and Physics 56~(11) (2006) 2289–2325.
\newblock \href {https://doi.org/10.1016/j.geomphys.2005.12.003} {\path{doi:10.1016/j.geomphys.2005.12.003}}.

\bibitem{sviratcheva2004physical}
K.~D. Sviratcheva, C.~Bahri, A.~I. Georgieva, J.~P. Draayer, Physical significance of    deformation and many-body interactions in nuclei, Physical Review Letters 93~(152501) (Oct. 2004).
\newblock \href {https://doi.org/10.1103/physrevlett.93.152501} {\path{doi:10.1103/physrevlett.93.152501}}.

\bibitem{edet2026controllable}
C.~O. Edet, E.~P. Inyang, O.~Abah, N.~Ali, Controllable diatomic molecular quantum thermodynamic machines, The European Physical Journal Plus 141 (2026) 282.
\newblock \href {https://doi.org/https://doi.org/10.1140/epjp/s13360-026-07434-w} {\path{doi:https://doi.org/10.1140/epjp/s13360-026-07434-w}}.

\bibitem{shannon1948mathematical}
C.~E. Shannon, A mathematical theory of communication, The Bell system technical journal 27~(3) (1948) 379--423.

\bibitem{moreira2025testing}
A.~R.~P. Moreira, A.~Bouzenada, O.~S. Oyun, F.~Ahmed, Testing shannon entropic measurement of the dirac oscillator under cosmic string geometry, Quantum Information Processing 24~(11) (2025) 374.
\newblock \href {https://doi.org/10.1007/s11128-025-05000-4} {\path{doi:10.1007/s11128-025-05000-4}}.

\bibitem{moreira2026quantuminformation}
Q.~R. D.~S. Moreira, L.~F. Ximenes, A.~R.~P. Moreira, J.~B.~R. Silva, Quantum information and thermodynamic features in position-dependent mass semiconductor heterostructures, Physica E 178 (2026) 116478.
\newblock \href {https://doi.org/10.1016/j.physe.2026.116478} {\path{doi:10.1016/j.physe.2026.116478}}.

\bibitem{fisher1925theory}
R.~A. Fisher, Theory of statistical estimation, Mathematical Proceedings of the Cambridge Philosophical Society 22~(5) (1925) 700–725.
\newblock \href {https://doi.org/https://doi.org/10.1017/s0305004100009580} {\path{doi:https://doi.org/10.1017/s0305004100009580}}.

\bibitem{frieden1992fisher}
B.~Frieden, Fisher information and uncertainty complementarity, Physics Letters A 169~(3) (1992) 123–130.
\newblock \href {https://doi.org/https://doi.org/10.1016/0375-9601(92)90581-6} {\path{doi:https://doi.org/10.1016/0375-9601(92)90581-6}}.

\bibitem{vignat2003analysis}
C.~Vignat, J.-F. Bercher, Analysis of signals in the fisher–shannon information plane, Physics Letters 312~(1-2) (2003) 27–33.
\newblock \href {https://doi.org/https://doi.org/10.1016/s0375-9601(03)00570-x} {\path{doi:https://doi.org/10.1016/s0375-9601(03)00570-x}}.

\bibitem{romera2004fisher}
E.~Romera, J.~S. Dehesa, The fisher–shannon information plane, an electron correlation tool, The Journal of Chemical Physics 120~(19) (2004) 8906–8912.
\newblock \href {https://doi.org/https://doi.org/10.1063/1.1697374} {\path{doi:https://doi.org/10.1063/1.1697374}}.

\bibitem{inyang2025quantum}
E.~P. Inyang, N.~Ali, R.~Endut, N.~Yusof, S.~Aljunid, M.~Romli, Quantum mechanical analysis of cesium dimer, titanium hydride, and titanium carbide: Vibrational spectra, expectation values, and information-theoretic measures, Chinese Journal of Physics 97 (Jul 2025).
\newblock \href {https://doi.org/https://doi.org/10.1016/j.cjph.2025.07.010} {\path{doi:https://doi.org/10.1016/j.cjph.2025.07.010}}.

\bibitem{sen2007fisher}
K.~D. Sen, J.~Antolín, J.~C. Angulo, Fisher-shannon analysis of ionization processes and isoelectronic series, Physical Review A 76~(3) (Sep 2007).
\newblock \href {https://doi.org/https://doi.org/10.1103/physreva.76.032502} {\path{doi:https://doi.org/10.1103/physreva.76.032502}}.

\bibitem{angulo2008atomic}
J.~C. Angulo, J.~Antolín, Atomic complexity measures in position and momentum spaces, The Journal of Chemical Physics 128~(16) (Apr 2008).
\newblock \href {https://doi.org/https://doi.org/10.1063/1.2907743} {\path{doi:https://doi.org/10.1063/1.2907743}}.

\bibitem{lopez2011statistical}
R.~López-Ruiz, J.~Sañudo, E.~Romera, X.~Calbet, Statistical complexity and fisher-shannon information: Applications, in: Statistical Complexity, Springer Netherlands, Dordrecht, 2011, p. 65–127.
\newblock \href {https://doi.org/https://doi.org/10.1007/978-90-481-3890-6_4} {\path{doi:https://doi.org/10.1007/978-90-481-3890-6_4}}.

\bibitem{jia2012equivalence}
C.-S. Jia, Y.-F. Diao, X.-J. Liu, P.-Q. Wang, G.-D. Zhang, Equivalence of the wei potential model and tietz potential model for diatomic molecules, The Journal of Chemical Physics | AIP Publishing 137~(014101) (2012).
\newblock \href {https://doi.org/https://doi.org/10.1063/1.4731340} {\path{doi:https://doi.org/10.1063/1.4731340}}.

\bibitem{beckner1975inequalities}
W.~Beckner, Inequalities in fourier analysis, Annals of Mathematics 102~(1) (1975) 159--182.

\bibitem{bialynicki1975uncertainty}
I.~Bia{\l}ynicki-Birula, J.~Mycielski, Uncertainty relations for information entropy in wave mechanics, Commun. Math. Phys 44~(2) (1975) 129--132.
\newblock \href {https://doi.org/10.1007/BF01608825} {\path{doi:10.1007/BF01608825}}.

\bibitem{romera2005fisher}
E.~Romera, P.~Sánchez-Moreno, J.~Dehesa, The fisher information of single-particle systems with a central potential, Chemical Physics Letters 414~(4-6) (2005) 468–472.
\newblock \href {https://doi.org/https://doi.org/10.1016/j.cplett.2005.08.032} {\path{doi:https://doi.org/10.1016/j.cplett.2005.08.032}}.

\bibitem{dembo1991information}
A.~Dembo, T.~Cover, J.~Thomas, Information theoretic inequalities, IEEE Transactions on Information Theory 37~(6) (1991) 1501–1518.
\newblock \href {https://doi.org/https://doi.org/10.1109/18.104312} {\path{doi:https://doi.org/10.1109/18.104312}}.

\bibitem{aquino2013shannon}
N.~Aquino, A.~Flores-Riveros, J.~Rivas-Silva, Shannon and fisher entropies for a hydrogen atom under soft spherical confinement, Physics Letters A 377~(34-36) (2013) 2062–2068.
\newblock \href {https://doi.org/https://doi.org/10.1016/j.physleta.2013.05.048} {\path{doi:https://doi.org/10.1016/j.physleta.2013.05.048}}.

\end{thebibliography}

\end{document}